# A Lax Pair for the Dynamics of DNA Modeled as a Shearable and Extensible Elastic Rod

Yaoming Shi(a), W.M. McClain(b), and John E. Hearst(a)

(a)Department of Chemistry, University of California, Berkeley, CA 94720-1460

(b)Department of Chemistry, Wayne State University, Detroit, MI 48202

Contact:  jehearst@cchem.berkeley.edu

**July 28, 2001**

## ABSTRACT

We introduce a spectral parameter into the geometrically exact Hamiltonian equations for the elastic rod in a way that creates a Lax pair. This assures integrability and permits application of the inverse scattering transform solution method. If the method can be carried through, the solution of the original problem is recovered by setting the spectral parameter to zero.

## I. INTRODUCTION

Consider a family of curves that are bendable and twistable (but not shearable or extendable), parametrized by arc length $s$ and time $t$, and governed by a set of partial differential equations. It has been recently shown by Doliwa and Santini[1] that that the equations are integrable, provided that three conditions hold:

(i)  the curves lie on (or in) the surface of an N-dimensional sphere;

(ii)  the total arclength of the curve is fixed; and

(iii)  the dynamics does not depend explicitly upon the radius of the sphere.

In this very general formulation there is no specific physical cause for the motion of the curve; it moves in any way that might be specified by its differential equations.

A systematic solution procedure is suggested by these results. Consider a kinematic problem in N-1 dimensions that is nonintegrable, in the sense that no inverse scattering procedure can be found that leads to its solution. Increase its parametrization by introducing a new independent variable    (the inverse of an N-dimensional sphere radius) plus new equation terms that require the curves to lie on the surface of this sphere. The expanded equations are now guaranteed integrable, and they lead to an inverse scattering procedure that will integrate them if it can be carried through. Now let  go to zero, flattening the sphere surface back to a Euclidian space of N-1 dimensions. Solutions in the infinitesimal neighborhood of any point on the sphere now become solutions to the original equations, which may not have been integrable (may not have been susceptible to an inverse scattering solution) on their own.

In this paper we extend these powerful ideas to the case of a curve that moves as a result of its own elasticity, known as the "elastic rod". In the geometrically exact Hamiltonian formulation, the dynamic behavior of the elastic rod[2] is governed by four vector equations. Two are pure kinematic equations that relate bending and twisting to local frame rotation, and shear and extension to local frame translation. The other two vector equations are dynamic equations that relate elastic forces, torques, and inertia terms to local frame rotation and translation. We show how these equations may be treated in an expanded scheme similar to that of Doliwa and Santini, except that constant arc length is no longer required. The original problem, of questionable integrability, consists of twelve scalar Hamiltonian equations, twelve scalar constitutive equations, and 24 dependent variables (each one a function of $s$ and $t$). We use the generator matrices of the Special Orthogonal Lie rotation group SO(4), under which the 4-dimensional sphere is fully symmetric, together with a real spectral parameter  , to construct two complex, 4-by-4 quantities $U$ and $V$, which form a Lax pair in the sense that they obey the Zakharov-Shabat form of the Lax equation

$$\partial_t U - \partial_s V + [U, V] = 0 \qquad (1.1)$$

This form insures integrability and also specifies a solution procedure via the inverse scattering transform[3]. If a solution can be actually constructed, the limit  0 gives the solution of the geometrically exact elastic rod problem.

This procedure includes as special cases the solution procedures for specialized reductions of the geometrically exact equations. For instance, the inextensible and unshearable Kirchhoff rod is included, as is the heavy top problem and the problem of rigid body motion in an ideal fluid.

At the end, we also present a connection between our system of PDEs with Myzakulov's recent unit spin description of soliton equations in (1+1) dimension[4].

## II. THE CONFIGURATION SPACE OF THE ELASTIC ROD IN 3D

We treat duplex DNA as a bendable, twistable, extensible, and shearable thin elastic rod. Here and elsewhere in this paper the terms "elastic rod" and "DNA" have the same meaning; so do "the centerline of the rod" and "the axis of DNA".

At a given time $t$ and at each point $s$ on the centerline $\mathbf{r}(s,t)$ of the rod, a localized Cartesian coordinate frame (or director frame), $\{\hat{\mathbf{d}}_1(s,t), \hat{\mathbf{d}}_2(s,t), \hat{\mathbf{d}}_3(s,t)\}$ is affixed with the unit vectors $\hat{\mathbf{d}}_1(s,t)$ and $\hat{\mathbf{d}}_2(s,t)$ in the direction of the principal axes of inertia tensor of the rod cross section. The third unit vector $\hat{\mathbf{d}}_3(s,t)$ is in the normal direction of the cross section. Because the shear is included, the unit vector $\hat{\mathbf{d}}_3(s,t)$ does not necessarily coincide with the tangent vector $(s,t)$ of the centerline of the elastic rod $\mathbf{r}(s,t)$.

Unit vectors $\hat{\mathbf{d}}_i(s,t)$ in the director frame (or body-fixed frame) are related to the unit vectors $\hat{\mathbf{a}}_i$ in the lab frame via an Euler rotation matrix according to $\hat{\mathbf{d}}_i(s,t) = ((s,t), (s,t), (s,t)) \hat{\mathbf{a}}_i$. Since $\mathbf{r} \in R^3$ and $\in SO(3)$, the configuration space of the elastic rod is $E^3 = R^3 \times SO(3)$.

The orientation of the local frame at $s + \Delta s$ is obtained by an infinitesimal rotation of the coordinate frame at $s$. The velocity of the rotation is the Darboux vector $\Omega$, with local components $\Omega = \Omega_i \hat{\mathbf{d}}_i$. Here and after, double occurrence of an index means summation over its range. Vectors $\hat{\mathbf{d}}_i$ change with $s$ by moving perpendicular to themselves and to the rotation axis $\Omega$ according to $\partial_s \hat{\mathbf{d}}_i = \Omega \times \hat{\mathbf{d}}_i$. At any time $t$ the relative position of the origin of the localized rod frame at $s + \Delta s$ is obtained by an infinitesimal translation $\Delta \mathbf{r}$ of the origin of the localized frame at $s$. The velocity of the translation is the tangent vector $\partial_s \mathbf{r} = \mathbf{t} = t_i \mathbf{d}_i$. The parameter $s$, usually chosen as the arclength parameter for the undeformed (or relaxed) elastic rod, is no longer the current arclength parameter for the deformed rod, $\tilde{s}(s,t)$, since there are deformations of shear and extension. The current arclength of the deformed rod, $\tilde{s}(s,t)$, is then given by $\tilde{s}(s,t) = \int^s |\mathbf{t}(\xi,t)| d\xi$.

The orientation of the local frame at time $t + \Delta t$ is obtained by an infinitesimal rotation of the coordinate frame at time $t$. The velocity of the rotation is the angular velocity vector $\omega$, with local frame components $\omega = \omega_i \hat{\mathbf{d}}_i$, obeying $\partial_t \hat{\mathbf{d}}_i = \omega \times \hat{\mathbf{d}}_i$. The relative position of the origin of the localized rod frame at $t + \Delta t$ is obtained by an infinitesimal translation of the origin of the localized frame at $t$. The velocity of the translation is the vector $\partial_t \mathbf{r} = \mathbf{v} = v_i \mathbf{d}_i$.

Three dependent variables are used for describing strains (or deformations) of bending$_1$ ($\kappa_1$), bending$_2$ ($\kappa_2$), and twisting ($\kappa_3$). Another three are used for describing the strains (or deformations) of shear$_1$ ($t_1$), shear$_2$ ($t_2$), and extension ($t_3$). Still another three are used for describing the linear velocity$_1$ ($v_1$), linear velocity$_2$ ($v_2$), and linear velocity$_3$ ($v_3$) for the translation of the centroid of the elastic rod cross section at position $s$. The last three are used for describing the angular velocity$_1$ ($\omega_1$), angular velocity$_2$ ($\omega_2$), and angular velocity$_3$ ($\omega_3$) for the rotation of the elastic rod cross section at position $s$.

At any time $t$ and a given position (say $s = s_1$) along the centerline, there is a cross section of the elastic rod upon which the internal forces are exerted. One side of the cross section ($s < s_1$) acts on the other side ($s > s_1$), and *vice versa*. The internal forces are resolved into a force $\mathbf{P}(s_1, t)$ and a torque $\mathbf{M}(s_1, t)$. At each cross section such a force and a torque may be found, giving rise to functions $\mathbf{P}(s, t)$ and $\mathbf{M}(s, t)$ describing system of stresses on the elastic rod.

Since each cross section has its mass and moment of inertia tensor, an angular momentum $\mathbf{m}(s, t)$ of the cross section and linear velocity $\mathbf{p}(s, t)$ of the center of the cross section can be naturally introduced.

From this point on, the letters $\omega, \Omega, M, P, \nu, \eta, m, p$ without subscripts will be understood as vectors, whether bold or not. The Simo-Marsden-Krishnaprasad (SMK) equations[5] governing these quantities, based on a geometrically exact Hamiltonian formulation of the problem, are

$$\Omega_t + \tfrac{1}{2} \omega \times \Omega = \omega_s + \tfrac{1}{2} \Omega \times \omega \tag{2.1a}$$

$$\nu_t + \omega \times \nu = \eta_s + \Omega \times \eta \tag{2.1b}$$

$$\partial_t p + \omega \times p = \partial_s P + \Omega \times P \tag{2.1c}$$

$$\partial_t m + \omega \times m + \nu \times p = \partial_s M + \Omega \times M + \eta \times P \tag{2.1d}$$

The constitutive relations are given by

$$P_i = \frac{\partial}{\partial \nu_i} E(\Omega, \nu, s) \tag{2.1e}$$

$$M_i = \frac{\partial}{\partial \Omega_i} E(\Omega, \nu, s) \tag{2.1f}$$

$$p_i = \frac{\partial}{\partial \eta_i} H(\eta, \omega, t) \tag{2.1g}$$

$$m_i = \frac{\partial}{\partial \dot{\Omega}_i} H(\ , \ ,t) \tag{2.1h}$$

where $i=1,2,3$ and $E(\ ,\ ,s)$ is the elastic energy function and $H(\ ,\ ,t)$ is the kinetic energy function.

We remark that the term $\omega \times p$ is not present in the original form of the SMK equations. This is because the kinetic energy $H(\ ,\ ,t)$ is chosen such that $p_i = \dot{\Omega}_i$ and $\omega \times p = 0$. We explicitly add this term here to show that SMK equations in (2.1) have the following exchange symmetry:

$$s \leftrightarrow t, \quad \Omega_i \leftrightarrow \omega_i, \quad \kappa_i \leftrightarrow \dot{\Omega}_i, \quad M_i \leftrightarrow m_i, \quad P_i \leftrightarrow p_i, \text{ and } E \leftrightarrow H.$$

After the equations have been solved for these dependent variables, one may construct the centerline $\mathbf{r}(s,t)$ and vectors $\hat{\mathbf{d}}_i(s,t)$ by the following procedure[6]:

(1) Solve $\partial_s \mathcal{A}(s,t) = \Omega(s,t) \wedge \mathcal{A}(s,t)$ and $\partial_t \mathcal{A}(s,t) = \omega(s,t) \wedge \mathcal{A}(s,t)$ for Euler matrix $\mathcal{A}(s,t)$, where $\Omega_{jk}(s,t) = \Omega_i(s,t)\varepsilon_{ijk}$ and $\omega_{jk}(s,t) = \omega_i(s,t)\varepsilon_{ijk}$.

(2) Calculate $\hat{\mathbf{d}}_i(s,t)$ using $\hat{\mathbf{d}}_i(s,t) = \mathcal{A}(s,t)\hat{\mathbf{a}}_i$.

(3) Calculate $\mathbf{r}(s,t) = \int_0^s \tau_i(x,t)\mathbf{d}_i(x,t)\,dx$ or $\mathbf{r}(s,t) = \int_0^t \nu_i(s,x)\mathbf{d}_i(s,x)\,dx$.

Even when these operations cannot be carried out analytically, numerical procedures give a very accurate picture of the dynamics.

## III. LAX PAIR FOR THE SMK EQUATIONS

Let $J_i$ and $K_i$ ($i=1,2,3$) be the generator matrices of Lie group $SO(4)$, given by

$$J_1 = \begin{pmatrix} 0 & 0 & 0 & 0 \\ 0 & 0 & -1 & 0 \\ 0 & 1 & 0 & 0 \\ 0 & 0 & 0 & 0 \end{pmatrix}, \quad J_2 = \begin{pmatrix} 0 & 0 & 1 & 0 \\ 0 & 0 & 0 & 0 \\ -1 & 0 & 0 & 0 \\ 0 & 0 & 0 & 0 \end{pmatrix}, \quad J_3 = \begin{pmatrix} 0 & -1 & 0 & 0 \\ 1 & 0 & 0 & 0 \\ 0 & 0 & 0 & 0 \\ 0 & 0 & 0 & 0 \end{pmatrix} \tag{3.1a}$$

$$K_1 = \begin{matrix} 0 & 0 & 0 & -1 \\ 0 & 0 & 0 & 0 \\ 0 & 0 & 0 & 0 \\ 1 & 0 & 0 & 0 \end{matrix}, \quad K_2 = \begin{matrix} 0 & 0 & 0 & 0 \\ 0 & 0 & 0 & -1 \\ 0 & 0 & 0 & 0 \\ 0 & 1 & 0 & 0 \end{matrix}, \quad K_3 = \begin{matrix} 0 & 0 & 0 & 0 \\ 0 & 0 & 0 & 0 \\ 0 & 0 & 0 & -1 \\ 0 & 0 & 1 & 0 \end{matrix} \quad (3.1b)$$

These generators satisfy the relations

$$[J_i, J_j] = \varepsilon_{ijk} J_k, \quad [J_i, K_j] = \varepsilon_{ijk} K_k, \quad [K_i, K_j] = \varepsilon_{ijk} J_k \quad (3.1c)$$

where $\varepsilon_{ijk}$ is the Levi-Civita symbol. Now consider the linear system

$$\psi_s(s, t, \lambda) = U(s, t, \lambda) \psi(s, t, \lambda) \quad (3.2a)$$

$$\psi_t(s, t, \lambda) = V(s, t, \lambda) \psi(s, t, \lambda) \quad (3.2b)$$

where $U$ and $V$ are defined as

$$U = A + \sqrt{-1}\, B \quad (3.3a)$$

$$V = C + \sqrt{-1}\, D \quad (3.3b)$$

and where $A(s,t,\lambda)$, $B(s,t,\lambda)$, $C(s,t,\lambda)$, and $D(s,t,\lambda)$ are given by

$$A = -\left(\Omega_i J_i + \lambda^2 \Omega_i K_i\right) \quad (3.3c)$$

$$B = -\left(p_i J_i + \lambda^2 m_i K_i\right) \quad (3.3d)$$

$$C = -\left(\omega_i J_i + \lambda^2 \omega_i K_i\right) \quad (3.3e)$$

$$D = -\left(P_i J_i + \lambda^2 M_i K_i\right) \quad (3.3f)$$

In system (3.3c-f), all symbols are real functions of $s$ and $t$ (except $\lambda$, the real spectral parameter). The integrability condition for system (3.2), $\psi_{ts} = \psi_{st}$, then leads to the Lax equation,

$$\partial_t U - \partial_s V + [U, V] = 0, \quad (3.4)$$

a defining property of a Lax pair. Note that $U$ and $V$ are 4-by-4 matrices. Left-multiplying or right-multiplying (3.4) by $J_i$ and $K_i$ ($i=1,2,3$) respectively, using (3.1c) to simplify the result, and separating the real part from the imaginary part, we obtain a system of PDEs in vector form:

$$\partial_t + \tfrac{1}{2} \times = \partial_s + \tfrac{1}{2} \times + {}^2(P \times p + {}^2 \times + {}^4 M \times m) \tag{3.5a}$$

$$\partial_t + \times = \partial_s + \times + {}^2(M \times p + P \times m) \tag{3.5b}$$

$$\partial_t p + \times p = \partial_s P + \times P + {}^4(m \times + \times M) \tag{3.5c}$$

$$\partial_t m + \times m + \times p = \partial_s M + \times M + \times P \tag{3.5d}$$

Since the system (3.5a-d) contains 12 scalar equations for 24 real dependent variables, $\alpha_i$, $\beta_i$, $\gamma_i$, $\delta_i$, $M_i$, $P_i$, $m_i$, $p_i$ ($i=1,2,3$), we have the freedom to pick twelve real "constitutive" relations. In elastic work one uses the twelve scalar equations implied by the four vector equations 2.1e-h, but in other problems one might use other relations, which we write symbolically as

$$E_A(\,,\,,M,P,\,,\,,m,p,s,t,\,)=0 \qquad (A=1,2,\ldots,12) \tag{3.5e}$$

System (3.5a-e) can describe a very large class of PDEs (24 dependent variables in (1+1) dimension). Some members of this class will be discussed in detail in Section IV, below.

In Appendix A, we give a more general formulation in which the 24 variables $\alpha_i$, $\beta_i$, $\gamma_i$, $\delta_i$, $M_i$, $P_i$, $m_i$, $p_i$ ($i=1,2,3$) are taken complex, and the Lax pair comprises two 8-by-8 matrices.

The system of PDEs in (3.5a-d) and its associated Lax pairs of (3.2-3.3) or (A.1-A.2) are the major results of this paper.

What is new about this Lax pair and the resulting system of 12 PDEs? The pure kinematic approach[7] for recasting a the nonlinear PDEs of a general curve into Lax representation focuses only on what we call the strain-velocity compatibility (integrability) relations for equations like

$$\partial_s = Q(\,,\,,s,t,\,), \qquad \partial_t = R(\,,\,,s,t,\,)$$

where and are strains and and are velocities. But we treat strain-velocity and stress-momentum on an equal footing, so our equations are

$$\partial_s \mathcal{E} = U(\mathcal{E}, \partial_s \mathcal{E}, p, m, s, t, \lambda), \qquad \partial_t \mathcal{E} = V(\mathcal{E}, \partial_s \mathcal{E}, P, M, s, t, \lambda)$$

where the new variables $P$ and $M$ are stresses (force and torque, respectively) and $p$ and $m$ are momenta (linear and angular, respectively). The Lax equations for the two cases look identical

$$\partial_t Q - \partial_s R + [Q, R] = 0 \quad versus \quad \partial_t U - \partial_s V + [U, V] = 0$$

and quantities $Q$, $R$, $U$, and $V$ are all 4-by-4 arrays, but $Q$ and $R$ are real, whereas our $U$ and $V$ are complex, and contain twice as many dependent variables.

We emphasize that the stress-momenta appear as naturally as the strain-velocities, and the resulting nonlinear PDEs are dynamic elastic equations rather than just kinematic equations.

Expanding the dependent variables $X_\mu = X_\mu(s, t, \lambda)$ and the constitutive relations $E_A(\mathcal{E}, \partial_s \mathcal{E}, M, P, \partial_s \mathcal{E}, \partial_t \mathcal{E}, m, p, s, t, \lambda)$ in Taylor series in $\lambda$,

$$X_i(s, t, \lambda) = \sum_{n=0}^\infty \lambda^n X_i^{(n)}(s, t) \text{ and } E_A(s, t, \lambda) = \sum_{n=0}^\infty \lambda^n E_{A\,i}^{(n)}(\ldots, s, t) \qquad (3.6)$$

and taking the limit $\lambda \to 0$, the leading terms in the expansions (3.5) are the SMK equations, 2.1a-d.

## IV. SPECIALIZATION OF THE SMK EQUATIONS

## Case 1. The Static Elastic Rod

Setting velocities and momenta $\partial_t \mathcal{E} = \partial_t \mathcal{E} = m = p = (0,0,0)^T$ in (2.1) and assuming that everything else is a function of independent variable $s$ only, the SMK equations (2.1) reduces to

$$\partial_s P + \kappa \times P = 0 \qquad (4.1a)$$

$$\partial_s M + \partial \times M + \partial \times P = 0 \qquad (4.1b)$$

$$P_i = \frac{\partial}{\partial \gamma_i} E(\kappa, \gamma, s) \qquad (4.1c)$$

$$M_i = \frac{\partial}{\partial \kappa_i} E(\kappa, \gamma, s) \qquad (4.1d)$$

System (4.1) describes the equilibrium configurations of an elastic rod with elastic energy function $E(\kappa, \gamma, s)$.

### Subcase 1.1  Static Elastic Rod with Linear Constitutive Relations

Let $E(\kappa, \gamma, s)$ of (4.1c) and (4.1d) be given by

$$E(\kappa, \gamma, s) = \tfrac{1}{2} A_{ij}\left(\kappa_i - \kappa_i^{(\text{intrinsic})}\right)\left(\kappa_j - \kappa_j^{(\text{intrinsic})}\right) + \tfrac{1}{2} C_{ij}\left(\gamma_i - \gamma_i^{(\text{intrinsic})}\right)\left(\gamma_j - \gamma_j^{(\text{intrinsic})}\right) + \tfrac{1}{2} B_{ij}\left[\left(\kappa_i - \kappa_i^{(\text{intrinsic})}\right)\left(\gamma_j - \gamma_j^{(\text{intrinsic})}\right) + \left(\gamma_i - \gamma_i^{(\text{intrinsic})}\right)\left(\kappa_j - \kappa_j^{(\text{intrinsic})}\right)\right]$$

$$(4.2)$$

where $A_{ij}(s)$ is the bending/twisting modulus, $C_{ij}(s)$ is the shear/extension modulus, and $B_{ij}(s)$ is coupling modulus between bending/twisting and shear/extension. The quantities $\kappa_i^{(\text{intrinsic})}(s)$ are the intrinsic bending and twisting of the unstressed rod, and the $\gamma_i^{(\text{intrinsic})}(s)$ are the intrinsic shear and extension. Then system (4.1) reduces to

$$\partial_s P + \partial \times P = 0 \qquad (4.3a)$$

$$\partial_s M + \partial \times M + \partial \times P = 0 \qquad (4.3b)$$

$$M_i = A_{ij}\left(\kappa_j - \kappa_j^{(\text{intrinsic})}\right) + B_{ij}\left(\gamma_i - \gamma_j^{(\text{intrinsic})}\right) \qquad (4.3c)$$

$$P_i = C_{ij}\left(\gamma_j - \gamma_j^{(\text{intrinsic})}\right) + B_{ij}\left(\kappa_j - \kappa_j^{(\text{intrinsic})}\right) \qquad (4.3d)$$

Further, when

(1) $B_{ij} = 0$ and

(2) $A_{ij}$ is a constant diagonal matrix with $A_1 = A_2$, and

(3) $C_{ij}$ is a constant diagonal matrix with $C_1 = C_2$, and

(4) $^{\text{(intrinsic)}} = (0,0,0)^T$, $^{\text{(intrinsic)}} = (0,0,1)^T$.

then system (4.3a,b) may be solved exactly in terms of elliptic functions[8].

### Subcase 1.2   Static Kirchhoff Elastic Rod, and the Heavy Top

In (4.3), let $_i - _i^{\text{(intrinsic)}} =$ , $B_{ij} = 0$, and $C_{ij} = _{ij} P_i /$ (no sum in $i$). Then $E(\ ,\ ,s)$ becomes

$$E(\ ,\ ,s) = \tfrac{1}{2} A_{ij} (_i - _i^{\text{(intrinsic)}})(_j - _j^{\text{(intrinsic)}}) + {}^{-1} P_i P_i \qquad (4.4a)$$

Take the limit $\ \to 0$ and absorb the last infinite term into the elastic energy $E(\ ,\ ,s)$. Then the elastic energy becomes

$$E(\ ,s) = \tfrac{1}{2} A_{ij} (_i - _i^{\text{(intrinsic)}})(_j - _j^{\text{(intrinsic)}}) \qquad (4.4b)$$

and the system (4.3) reduces to

$$_s P + \ \times P = 0 \qquad (4.5a)$$

$$_s M + \ \times M + \ \times P = 0 \qquad (4.5b)$$

$$M_i = A_{ij}(_j - _j^{\text{(intrinsic)}}) \qquad (4.5c)$$

If $s$ is arc length (as assumed earlier), then system (4.5) describes the equilibrium configuration of the unshearable, inextensible Kirchhoff elastic rod. But if $s$ is understood as time, this system describes the dynamics of the heavy top.

There are two known integrable cases for the heavy top system (4.5) with $^{\text{(intrinsic)}} = (0,0,0)^T$, $= (0,0,1)^T$:

(a) Lagrange Top[9]: $A_{ij}$ is a constant diagonal matrix with $A_1 = A_2$

(b) Kowalewski Top[10]: the same, but with $A_3 = A_1 = 2A_2$.

## Case 2, Rigid Body Motion In Ideal Fluid

Setting $M = P = \ = \ = (0,0,0)^T$ in (2.1) and assuming that everything else is function of time $t$ only, then the SMK equations (2.1) reduce to

$$\partial_t p + \ \times p = 0 \tag{5.1a}$$

$$\partial_t m + \ \times m + \ \times p = 0 \tag{5.1b}$$

$$\Omega_i = \frac{\partial}{\partial p_i} H(m, p, t) \tag{5.1c}$$

$$\Omega_i = \frac{\partial}{\partial m_i} H(m, p, t) \tag{5.1d}$$

Let $H(m, p, t)$ of (5.1c) and (5.1d) be given by

$$H(m, p, t) = \tfrac{1}{2} a_{ij}\left(m_i - m_i^{(0)}\right)\left(m_j - m_j^{(0)}\right) + \tfrac{1}{2} c_{ij}\left(p_i - p_i^{(0)}\right)\left(p_j - p_j^{(0)}\right)$$
$$+ \tfrac{1}{2} b_{ij}\left[\left(m_i - m_i^{(0)}\right)\left(p_j - p_j^{(0)}\right) + \left(m_j - m_j^{(0)}\right)\left(p_i - p_i^{(0)}\right)\right] \tag{5.2}$$

where $a_\mu$, $b_\mu$, $c_\mu$, $m_\mu^{(0)}$, $p_\mu^{(0)}$ are constants. Then system (5.1) reduces to the Kirchhoff equations, which describe a finite rigid body moving in an ideal incompressible fluid:

$$\partial_t p + \ \times p = 0 \tag{5.3a}$$

$$\partial_t m + \ \times m + \ \times p = 0 \tag{5.3b}$$

$$\Omega_i = c_{ij}\left(p_j - p_j^{(0)}\right) + b_{ij}\left(m_j - m_j^{(0)}\right) \tag{5.3c}$$

$$\Omega_i = a_{ij}\left(m_j - m_j^{(0)}\right) + b_{ij}\left(p_j - p_j^{(0)}\right) \tag{5.3d}$$

There exist three known non-trivial integrable cases:

(1) Clebsch case [11]: $m_i^{(0)} = p_i^{(0)} = 0$, $a_{ij} = a_{i\ ij}$, $b_{ij} = 0$, $c_{ij} = c_{i\ ij}$, and $a_i$, $c_i$ are constants satisfying $a_1^{-1}(c_2 - c_3) + a_2^{-1}(c_3 - c_1) + a_3^{-1}(c_1 - c_2) = 0$,

(2) Steklov case [12]: $m_i^{(0)} = p_i^{(0)} = 0$, $a_{ij} = a_{i\ ij}$, $b_{ij} = 0$, $c_{ij} = c_{i\ ij}$, and $a_i$, $c_i$ are constants satisfying $b_i = (a_1 a_2 a_3) a_i^{-1} +$ , $c_1 = {}^2 a_1 (a_2 - a_3)^2 +$ , $c_2 = {}^2 a_2 (a_3 - a_1)^2 +$ , $c_3 = {}^2 a_3 (a_1 - a_2)^2 +$ , and , , and are constants.

(3) Chaplygin case [13].

## Case 3, Kirchhoff Elastic Rod Motion

In this case, there is no shear or extension, so $= (0,0,1)^T$. Eqs. (2.1a) and (2.1b) are derived from the integrability conditions

$$\partial_s \partial_t \hat{\mathbf{d}}_i = \partial_t \partial_s \hat{\mathbf{d}}_i. \tag{6.1a}$$

$$\partial_s \partial_t \mathbf{r} = \partial_t \partial_s \mathbf{r} \tag{6.1b}$$

Since $= (0,0,1)^T$ we have $\partial_s \mathbf{r} = \hat{\mathbf{d}}_3$. Thus we can differentiate both sides of (6.1b) with respect to $s$ and obtain:

$$\partial_s \partial_t \hat{\mathbf{d}}_3 = \partial_t \partial_s \hat{\mathbf{d}}_3 \tag{6.1c}$$

The resulting Equation (6.1c) is actually the third component part of (6.1a).

SMK equations in (2.1) reduce to:

$$\partial_t + \frac{1}{2} \times = \partial_s + \frac{1}{2} \times \tag{6.2a}$$

$$\partial_t p + \times p = \partial_s P + \times P \tag{6.2c}$$

$$\partial_t m + \times m + \times p = \partial_s M + \times M + \times P \tag{6.2d}$$

System (6.2) describes the dynamics of Kirchhoff elastic rod if we pick

$$M_i = A_{ij}(\ _j - \ _j^{(\text{intrinsic})}) \tag{6.2e}$$

$$m_i = a_{ij}\,\omega_j \tag{6.2f}$$

$$p_i = \rho\,v_i \tag{6.2g}$$

where $a_{ij}$ is the moment of inertia tensor for the cross section of the elastic rod and $\rho$ is the linear density of mass.

## Case 4, Elastic Rod Moving in a Plane

Setting $v_1 = v_2 = m_1 = m_2 = \omega_3 = p_3 = 0$ and $\omega_1 = \omega_2 = M_1 = M_2 = \kappa_3 = P_3 = 0$ in (2.1) then SMK equations (2.1) reduce to

$$\partial_t v_3 - \partial_s v_3 = 0 \tag{7.1a}$$

$$\partial_t p_1 - \partial_s P_1 - p_2 \kappa_3 + \tau_3 P_2 = 0 \tag{7.1b}$$

$$\partial_t p_2 - \partial_s P_2 + p_1 \kappa_3 - \tau_3 P_1 = 0 \tag{7.1c}$$

$$\partial_t \kappa_1 - \partial_s \tau_1 - \kappa_2 v_3 + \tau_3 v_2 = 0 \tag{7.1d}$$

$$\partial_t \kappa_2 - \partial_s p_2 + \kappa_1 v_3 - \tau_3 v_1 = 0 \tag{7.1e}$$

$$\partial_t m_3 - \partial_s M_3 - p_1 v_2 + \kappa_1 p_2 - \kappa_1 P_2 + P_1 \kappa_2 = 0 \tag{7.1f}$$

System (7.1a – 7.1f) contains six scalar equations for twelve dependent variables. The six constitutive relations can be expressed as

$$P_1 = -\partial_{\kappa_1} E(\kappa_3, \kappa_1, \kappa_2, s) \tag{7.1g}$$

$$P_2 = -\partial_{\kappa_2} E(\kappa_3, \kappa_1, \kappa_2, s) \tag{7.1h}$$

$$M_3 = -\partial_{\kappa_3} E(\kappa_3, \kappa_1, \kappa_2, s) \tag{7.1i}$$

$$p_1 = -\partial_{v_1} H(v_3, v_1, v_2, t) \tag{7.1j}$$

$$p_2 = -\partial_{v_2} H(v_3, v_1, v_2, t) \tag{7.1k}$$

$$m_3 = \frac{\partial}{\partial \omega_3} H(\omega_3, \omega_1, \omega_2, t) \tag{7.11}$$

In system (7.1) dependent variable $\kappa_3$ is used for describing bending, $\kappa_1$ for shear, $\kappa_2$ for extension, $\omega_1$ for linear velocity$_1$, $\omega_2$ for linear velocity$_2$, and $\omega_3$ for angular velocity. $E(\kappa_3, \kappa_1, \kappa_2, s)$ stands for the elastic energy and $H(\omega_3, \omega_1, \omega_2, t)$ for the kinetic energy.

## Case 5, Moving Space Curve (3D)

Doliwa and Santini[14] have shown that for a space curve moving on a real 3-dimensional spherical surface $S^3$ with radius $\varepsilon^{-2}$, the Serret-Frenet frame $\{\hat{\mathbf{n}}(\tilde{s},t,\varepsilon), \hat{\mathbf{b}}(\tilde{s},t,\varepsilon), \hat{\mathbf{t}}(\tilde{s},t,\varepsilon), \hat{\mathbf{r}}(\tilde{s},t,\varepsilon)\}$ satisfies the relation

$$\partial_{\tilde{s}} \begin{pmatrix} \hat{\mathbf{b}} \\ \hat{\mathbf{n}} \\ \hat{\mathbf{t}} \\ \hat{\mathbf{r}} \end{pmatrix} = \begin{pmatrix} 0 & -\tau & 0 & 0 \\ \tau & 0 & -\kappa & 0 \\ 0 & \kappa & 0 & -\varepsilon^2 \\ 0 & 0 & \varepsilon^2 & 0 \end{pmatrix} \begin{pmatrix} \hat{\mathbf{b}} \\ \hat{\mathbf{n}} \\ \hat{\mathbf{t}} \\ \hat{\mathbf{r}} \end{pmatrix} \tag{8.1a}$$

where $\tilde{s}$ is the arclength, the unit radius vector is $\hat{\mathbf{r}}(\tilde{s},t,\varepsilon) = \varepsilon^{-2} \mathbf{r}(\tilde{s},t,\varepsilon)$, the curvature is $\kappa = \kappa(\tilde{s},t,\varepsilon)$, and the geometric torsion is $\tau = \tau(\tilde{s},t,\varepsilon)$.

The change of the Serret-Frenet frame with respect to time is assumed to be

$$\partial_t \begin{pmatrix} \hat{\mathbf{b}} \\ \hat{\mathbf{n}} \\ \hat{\mathbf{t}} \\ \hat{\mathbf{r}} \end{pmatrix} = \begin{pmatrix} 0 & \alpha_3 & -\alpha_2 & -\varepsilon^2 \beta_1 \\ -\alpha_3 & 0 & -\alpha_1 & -\varepsilon^2 \beta_2 \\ \alpha_2 & \alpha_1 & 0 & -\varepsilon^2 \beta_3 \\ \beta_1 & \beta_2 & \beta_3 & 0 \end{pmatrix} \begin{pmatrix} \hat{\mathbf{b}} \\ \hat{\mathbf{n}} \\ \hat{\mathbf{t}} \\ \hat{\mathbf{r}} \end{pmatrix} \tag{8.1b}$$

where $t$ is time and $X = X(\tilde{s},t,\varepsilon)$, with $X = \alpha_i$, and $\beta_i$, respectively.

The first two kinematic vector equations (2.1a,b) in the SMK equations (2.1) are derived from the following Serret-Frenet equations for an elastic rod moving on the 3-dimensional surface of a 4-dimensional sphere with radius $\varepsilon^{-2}$; namely,

$$\partial_s \begin{pmatrix} \hat{\mathbf{d}}_1 \\ \hat{\mathbf{d}}_2 \\ \hat{\mathbf{d}}_3 \\ \hat{\mathbf{r}} \end{pmatrix} = \begin{pmatrix} 0 & \kappa_3 & -\kappa_2 & -\varepsilon^2 \tau_1 \\ -\kappa_3 & 0 & \kappa_1 & -\varepsilon^2 \tau_2 \\ \kappa_2 & -\kappa_1 & 0 & -\varepsilon^2 \tau_3 \\ \tau_1 & \tau_2 & \tau_3 & 0 \end{pmatrix} \begin{pmatrix} \hat{\mathbf{d}}_1 \\ \hat{\mathbf{d}}_2 \\ \hat{\mathbf{d}}_3 \\ \hat{\mathbf{r}} \end{pmatrix} \quad (8.2a)$$

$$\partial_t \begin{pmatrix} \hat{\mathbf{d}}_1 \\ \hat{\mathbf{d}}_2 \\ \hat{\mathbf{d}}_3 \\ \hat{\mathbf{r}} \end{pmatrix} = \begin{pmatrix} 0 & \omega_3 & -\omega_2 & -\varepsilon^2 \sigma_1 \\ -\omega_3 & 0 & \omega_1 & -\varepsilon^2 \sigma_2 \\ \omega_2 & -\omega_1 & 0 & -\varepsilon^2 \sigma_3 \\ \sigma_1 & \sigma_2 & \sigma_3 & 0 \end{pmatrix} \begin{pmatrix} \hat{\mathbf{d}}_1 \\ \hat{\mathbf{d}}_2 \\ \hat{\mathbf{d}}_3 \\ \hat{\mathbf{r}} \end{pmatrix} \quad (8.2b)$$

If we assume zero elastic energy $E(\kappa, \tau, s)$ and zero kinetic energy $H(\omega, \sigma, t)$ in the constitutive relations (2.1e-h), then $P = M = p = m = (0,0,0)^T$. Consequently the last two dynamic vector equations (2.1c,d) in the SMK equations (2.1) for force balance and torque balance become zero identically.

Comparing (8.2a,b) with (8.1a,b) and identifying $\tilde{s}(s,t) = \int^s \tau_3(x,t)\,dx$, $\partial_s = (\tau_3)^{-1} \partial_{\tilde{s}}$, $\hat{\mathbf{d}}_3 = \hat{\mathbf{t}}$, $\hat{\mathbf{d}}_2 = \hat{\mathbf{n}}$, $\hat{\mathbf{d}}_1 = \hat{\mathbf{b}}$, we realize that a space curve is an elastic rod specialized in the following way:

$$\kappa = \kappa_1/\tau_3, \quad \tau = -\kappa_3/\tau_3. \quad (8.3a)$$

$$\kappa_2 = \tau_1 = \tau_2 = 0, \quad (8.3b)$$

Thus a space curve is a special case of elastic rod with no shear deformation ($\epsilon_1 = \epsilon_2 = 0$), no bending deformation in $\hat{\mathbf{d}}_2 = \hat{\mathbf{n}}$ direction ($\kappa_2 = 0$), zero elastic energy function $E(\kappa, \tau, s)$, and zero kinetic energy function $H(\omega, \nu, t)$.

This is the key link connecting the SMK equations (2.1), via the moving curve problem (8.1a,b), with most well-known integrable systems in dimension (1+1) (KdV, mKdV, sine-Gordon, nonlinear Schroedinger, Heisenberg ferromagnetism )[15].

There also exists another tetrad frame $(\hat{\mathbf{e}}_1(s,t,\lambda), \hat{\mathbf{e}}_2(s,t,\lambda), \hat{\mathbf{e}}_3(s,t,\lambda), \hat{\mathbf{e}}_4(s,t,\lambda))$ which satisfies the following relations:

$$\partial_s \begin{pmatrix} \hat{\mathbf{e}}_1 \\ \hat{\mathbf{e}}_2 \\ \hat{\mathbf{e}}_3 \\ \hat{\mathbf{e}}_4 \end{pmatrix} = \begin{pmatrix} 0 & p_3 & -p_2 & -\lambda^2 m_1 \\ -p_3 & 0 & -p_1 & -\lambda^2 m_2 \\ -p_2 & p_1 & 0 & -\lambda^2 m_3 \\ \lambda^2 m_1 & \lambda^2 m_2 & \lambda^2 m_3 & 0 \end{pmatrix} \begin{pmatrix} \hat{\mathbf{e}}_1 \\ \hat{\mathbf{e}}_2 \\ \hat{\mathbf{e}}_3 \\ \hat{\mathbf{e}}_4 \end{pmatrix} \quad (8.4a)$$

$$\partial_t \begin{pmatrix} \hat{\mathbf{e}}_1 \\ \hat{\mathbf{e}}_2 \\ \hat{\mathbf{e}}_3 \\ \hat{\mathbf{e}}_4 \end{pmatrix} = \begin{pmatrix} 0 & P_3 & -P_2 & -\lambda^2 M_1 \\ -P_3 & 0 & -P_1 & -\lambda^2 M_2 \\ -P_2 & P_1 & 0 & -\lambda^2 M_3 \\ \lambda^2 M_1 & \lambda^2 M_2 & \lambda^2 M_3 & 0 \end{pmatrix} \begin{pmatrix} \hat{\mathbf{e}}_1 \\ \hat{\mathbf{e}}_2 \\ \hat{\mathbf{e}}_3 \\ \hat{\mathbf{e}}_4 \end{pmatrix} \quad (8.4b)$$

Since the elements in 4-by-4 matrices in (8.2a,b) are related to the elements of the 4-by-4 matrices in (8.4a,b) via the constitutive relations like (2.1e-h), we may say that the frame $\{\hat{\mathbf{e}}_1, \hat{\mathbf{e}}_2, \hat{\mathbf{e}}_3, \hat{\mathbf{e}}_4\}$ is dual to the Serret-Frenet frame $\{\hat{\mathbf{b}}, \hat{\mathbf{n}}, \hat{\mathbf{t}}, \hat{\mathbf{r}}\}$.

## V. RELATION TO SPIN DESCRIPTION OF SOLITON EQUATIONS

We may rewrite the first component equation in (8.2b) as

$$\partial_t \hat{\mathbf{S}} = \sum_{J=2}^{4} b_J \hat{\mathbf{d}}_J, \qquad \hat{\mathbf{S}} \equiv \hat{\mathbf{d}}_1 \qquad (9.1)$$

This is a basic equation in Myrzakulov's unit spin description of the integrable and nonintegrable PDEs [16].

Eqs.(8.2a,b) tell us that it is better to consider the motion of not just unit vector $\hat{\mathbf{S}} \equiv \hat{\mathbf{d}}_1$, but all unit vectors $\hat{\mathbf{d}}_J$ ($J = 1,2,3,4$) together in (1+1) dimension. We conjecture that any system of integrable or nonintegrable PDEs in (1+1) dimension derived from Eq.(9.1) might also be derived from a system of PDEs in (3.5) with a Lax pair of (3.3-3.4) or (A.2-A.3).

If we use the 8-by-8 Lax pair as shown in Appendix A and choose the normalization factor properly for $\Psi$ in (A.1), then we can rewrite (A.1) as

$$\partial_s \hat{\mathbf{f}}_I = X_{IJ} \hat{\mathbf{f}}_J \qquad (I = 1,2,...,8) \qquad (9.2a)$$

$$\partial_t \hat{\mathbf{f}}_I = Y_{IJ} \hat{\mathbf{f}}_J, \qquad (I = 1,2,...,8) \qquad (9.2b)$$

where $\left(\hat{\mathbf{f}}_1, \hat{\mathbf{f}}_2, ..., \hat{\mathbf{f}}_8\right)^T = \Psi^T$, $\hat{\mathbf{f}}_I \cdot \hat{\mathbf{f}}_J = \delta_{IJ}$, and $X$ and $Y$ are the 8-by-8 matrices defined in (A.2a) and (A.2b) respectively.

Matrices $X$ and $Y$ have the following symmetry properties:

$$X_{IJ} = X_{JI}, \quad Y_{IJ} = Y_{JI} \quad \text{(if } I,J-4 = 1, 2, 3, 4 \text{ or } I-4, J = 1,2,3,4\text{)} \quad (9.3a)$$

$$X_{IJ} = -X_{JI}, Y_{IJ} = -Y_{JI} \quad \text{(if } I,J = 1,2,3,4 \text{ or } I,J = 5,6,7,8\text{)} \quad (9.3b)$$

Since matrix $X$ is not antisymmetric, Eq.(9.2a) cannot be considered as the Serret-Frenet equations for an elastic rod moving on 7-sphere ($S^7$) with radius $||\ ||^{-2} = {}^{-2}$ imbedded in $R^8$.

Because the diagonal matrix elements of $Y$ are all zero, we may rewrite the first component equation in (9.2b) as

$$\partial_t \hat{\mathbf{S}} = \sum_{J=2}^{8} a_J \hat{\mathbf{f}}_J, \quad a_J \quad Y_{1J}, \quad \hat{\mathbf{S}} \quad \hat{\mathbf{f}}_1 \quad (9.4)$$

This is the analog of Eq.(9.1) in 8D [17]. Again Eqs.(8.6a,b) tell us that it is better to consider the motion of not just unit vector $\hat{\mathbf{S}} \quad \hat{\mathbf{f}}_1$, but all unit vectors $\hat{\mathbf{f}}_J$ ($J = 1,2,...,7,8$) together in (1+1) dimension.

## VI. CONCLUSIONS

(1) We have found a Lax pair with a corresponding spectral parameter for a system of 12 scalar PDEs and 12 scalar constitutive relations governing 24 dependent variables in (1+1) dimension.

(2) When the spectral parameter goes to zero, this system of PDEs reduces to the SMK equations that describe the dynamics of DNA modeled as a shearable and extensible elastic rod.

(3) When three dependent variables are set to be constants, the SMK equations reduce to a system of 9 scalar PDEs with 9 scalar constitutive relations in 18 dependent variables. This system describes the dynamics of DNA modeled as an unshearable and inextensible elastic rod (Kirchhoff elastic rod).

(4) When the SMK equations are assumed to be independent of $t$ or $s$, they reduce to a set of 6 ODEs and 3 constitutive relations for 9 dependent variables describing the motion of a heavy top, or the motion of a rigid body in an ideal fluid.

It is beyond the scope of this paper to show in a mathematically rigorous fashion that each step in the reduction will (or will not) preserve the integrable properties of the PDEs (in the sense of Liouville, Painlevé, Lax, or Inverse Scattering Transformation (IST)).

## AKNOWLEDGEMENT

Y.S. would like to thank Zixiang Zhou of Fudan University, Shanghai, P.R. China for helpful discussion.

# APPENDIX A

If we assume that 24 dependent variables $\alpha_i$, $\beta_i$, $\gamma_i$, $\delta_i$, $M_i$, $P_i$, $m_i$, $p_i$ ($i=1,2,3$) are complex and the parameter $\lambda$ is also complex, then we may consider the following linear system:

$$\psi_s(s,t,\lambda) = X(s,t,\lambda)\psi(s,t,\lambda) \tag{A.1a}$$

$$\psi_t(s,t,\lambda) = Y(s,t,\lambda)\psi(s,t,\lambda) \tag{A.1b}$$

where $X$ and $Y$ in the Lax pair $X$-$Y$ are defined as

$$X = \begin{pmatrix} A & -B \\ B & A \end{pmatrix} \tag{A.2a}$$

$$Y = \begin{pmatrix} C & -D \\ D & C \end{pmatrix} \tag{A.2b}$$

and where $A(s,t,\lambda)$, $B(s,t,\lambda)$, $C(s,t,\lambda)$, and $D(s,t,\lambda)$ are given by

$$A = -\alpha_i J_i + \lambda^2 \beta_i K_i \tag{A.2c}$$

$$B = \left(-p_i J_i + \lambda^2 m_i K_i\right) \tag{A.2d}$$

$$C = -\gamma_i J_i + \lambda^2 \delta_i K_i \tag{A.2e}$$

$$D = \left(-P_i J_i + \lambda^2 M_i K_i\right) \tag{A.2f}$$

The integrability condition for system (A.1), $\psi_{ts} = \psi_{st}$, leads to:

$$\partial_t X - \partial_s Y + [X, Y] = 0 \tag{A.3}$$

Taking each 4-by-4 block in (A.3) and left- or right-multiplying it by $J_i$ and $K_i$ ($i=1,2,3$) respectively, using (3.1) to simplify the result, we obtain a set of PDEs in vector form:

$$\partial_t \bm{\kappa} + \tfrac{1}{2}\bm{\kappa} \times \bm{\kappa} = \partial_s \bm{\Omega} + \tfrac{1}{2}\bm{\Omega} \times \bm{\Omega} - \lambda^2(P \times p - \lambda^2 \bm{\kappa} \times \bm{\Omega} + \lambda^4 M \times m) \tag{A.4a}$$

$$\partial_t \bm{\Omega} + \bm{\kappa} \times \bm{\Omega} = \partial_s \bm{\kappa} + \bm{\Omega} \times \bm{\kappa} - \lambda^2(M \times p + P \times m) \tag{A.4b}$$

$$\partial_t p + \bm{\kappa} \times p = \partial_s P + \bm{\Omega} \times P + \lambda^4(m \times \bm{\kappa} + \bm{\Omega} \times M) \tag{A.4c}$$

$$\partial_t m + \bm{\kappa} \times m + \bm{\Omega} \times p = \partial_s M + \bm{\Omega} \times M + \bm{\kappa} \times P \tag{A.4d}$$

Assuming that $\lambda$ is pure imaginary, system (A.4) reduces to system (3.5).